\documentstyle[twoside,fleqn,espcrc2]{article}

\newcommand{\AmS}{{\protect\the\textfont2
  A\kern-.1667em\lower.5ex\hbox{M}\kern-.125emS}}

\title{ A solution to the puzzles of CP violation, neutrino oscillation, 
fermion masses and mixings in an SUSY GUT model 
with small $\tan \beta$}

\author{K.C. Chou \address{Chinese Academy of Sciences, Beijing 100864, China}
        and 
        Yue-Liang Wu\address{Department of Physics, Ohio State University, 
        174 W. 18th Ave., Columbus, OH 43210, USA}%
        \thanks{Supported in part by the US Department of Energy Grant \#
DOE/ER/01545-675.\ Permanent address: Institute of Theoretical Physics, Chinese
Academy of Sciences, Beijing 100080, China }}
       
\begin{document}

\begin{abstract}
CP violation, fermion masses and mixing angles including that of neutrinos are
studied in an SUSY SO(10)$\times \Delta(48) \times$ U(1) model with small 
$\tan\beta$. It is amazing that the model can provide a successful
prediction on twenty three observables by only using four parameters.
The renormalization group (RG) effects containing those above the GUT scale
are considered. Fifteen relations among the low energy parameters are found
with nine of them free from  RG modifications. 
 They could be tested directly by low energy experiments.
\end{abstract}

\maketitle

The standard model (SM) is a great success. But it cannot be a fundamental
theory. Eighteen phenomenological parameters have been introduced to 
describe the real world, all of unkown origin.
The mass spectrum and the mixing angles observed 
remind us that we are in a stage similar to that of  atomic 
spectroscopy before Balmer. 
In this talk, we shall present an interesting model 
based on the symmetry group SUSY $SO(10)\times \Delta(48)\times U(1)$
with small values of $\tan \beta \sim 1 $ which 
is of phenomenological interest in testing the Higgs 
sector in the minimum supersymmetric standard model (MSSM)
at   Colliders\cite{GKE}. For a detailed analysis see ref.
\cite{CHOUWU}. The dihedral group $\Delta (48)$, 
a subgroup of SU(3), is taken as the family group.  
$U(1)$ is family-independent and is introduced to distinguish 
various fields which belong to the same 
representations of $SO(10)\times \Delta (48)$.
The irreducible representations of $\Delta (48)$ consisting of five triplets 
and three singlets  are found to be sufficient to build an interesting
texture structure for fermion mass matrices.  The 
symmetry $\Delta(48)\times U(1)$  naturally 
ensures the texture structure with zeros for Yukawa coupling matrices, 
while the coupling coefficients of the resulting interaction terms in the
superpotential are unconstrainted by this symmetry. 
To reduce the possible free parameters, the universality of Yukawa coupling 
constants in the superpotential is assumed, i.e., 
all the coupling coefficients are assumed to be equal and have the same 
origins from perhaps a more fundamental theory. 

 Choosing the structure of the physical vacuum carefully, 
the Yukawa coupling matrices which 
determine the masses and mixings of all quarks and leptons are given at the GUT
scale by
\[ \Gamma_{u}^{G} = \lambda_{u} 
\left( \begin{array}{ccc} 
0  &  \frac{3}{2}z'_{u} \epsilon_{P}^{2} &   0   \\
\frac{3}{2}z_{u} \epsilon_{P}^{2} &  - 3 y_{u} 
\epsilon_{G}^{2} e^{i\phi}  
& -\frac{\sqrt{3}}{2}x_{u}\epsilon_{G}^{2}  \\
0  &  - \frac{\sqrt{3}}{2}x_{u}\epsilon_{G}^{2}  &  w_{u} 
\end{array} \right) \]
\[ \Gamma_{f}^{G} = \lambda_{f} 
\left( \begin{array}{ccc} 
0  &  -\frac{3}{2}z'_{f} \epsilon_{P}^{2} &   0   \\
-\frac{3}{2}z_{f} \epsilon_{P}^{2} &  3 y_{f} 
\epsilon_{G}^{2} e^{i\phi}  
& -\frac{1}{2}x_{f}\epsilon_{G}^{2}  \\
0  &  -\frac{1}{2}x_{f}\epsilon_{G}^{2}  &  w_{f} 
\end{array} \right)  \]
\[ \Gamma_{\nu}^{G} = \lambda_{\nu}
\left( \begin{array}{ccc} 
0  &  -\frac{15}{2}z'_{\nu} \epsilon_{P}^{2} &   0   \\
-\frac{15}{2}z_{\nu} \epsilon_{P}^{2} &  15 y_{\nu} 
\epsilon_{G}^{2} e^{i\phi}  
& -\frac{1}{2}x_{\nu}\epsilon_{G}^{2}  \\
0  &  -\frac{1}{2}x_{\nu}\epsilon_{G}^{2}  &  w_{\nu} 
\end{array} \right) \]
with $\lambda_{u} = 2\lambda_{H}/3$,
$\lambda_{f}=\lambda_{u}(-1)^{n+1}/3^{n}$ ($f=d,e$) and $\lambda_{\nu} =
\lambda_{f}/5^{n+1}$.
Here the integer $n$ reflects the possible
choice of heavy fermion fields above the GUT scale.  $n=4$ is found to be the
best choice in this set of models for 
a consistent prediction on top and charm quark masses.
This is because for $n > 4$, the resulting value of $\tan \beta$ becomes too
small, as a consequence, the predicted top quark mass will be below the present
experimental lower limit. For $ n < 4$, the values of $\tan \beta$ will become
larger, the resulting charm quark mass will be above the present upper bound. 
$\lambda_{H}=\lambda_{H}^{0}r_{3}$, $\epsilon_{G}\equiv (\frac{v_{5}}{v_{10}})
\sqrt{\frac{r_{2}}{r_{3}}}$ and $\epsilon_{P}\equiv
(\frac{v_{5}}{\bar{M}_{P}})\sqrt{\frac{r_{1}}{r_{3}}}$ are three parameters. 
Where $\lambda_{H}^{0}$ is a universal 
coupling constant expected to be of order one, $r_{1}$, $r_{2}$ and $r_{3}$ 
denote the ratios of the coupling constants of the superpotential at 
the GUT scale for the textures `12', `22' (`32') and `33' respectively. 
They represent the possible renormalization group (RG) effects 
running from the scale $\bar{M}_{P}$ to the GUT scale. Note that the RG effects
for the  textures `22' and `32' are considered to be the same since they are
generated from a similar superpotential structure after integrating out the
heavy fermions and concern the fields which belong to 
the same representations of the symmetry group, this can be explicitly seen 
from their effective operators $W_{22}$ and $W_{32}$ given below.
$\bar{M}_{P}$, $v_{10}$, and 
$v_{5}$ being the vacuum expectation values (VEVs) 
for $U(1)\times \Delta(48)$, SO(10) and SU(5) 
symmetry breaking respectively. $\phi$ is the physical CP phase 
arising from the VEVs. The assumption of maximum CP violation implies that 
$\phi = \pi/2$. 
$x_{f}$, $y_{f}$, $z_{f}$, and $w_{f}$ $(f = u, d, e, \nu)$ 
are the Clebsch factors of $SO(10)$ determined by the 
directions of symmetry breaking of the adjoints {\bf 45}'s. 
The Clebsch factors associated with the symmetry breaking directions can be 
easily read off from the U(1) hypercharges of the adjoints {\bf 45}'s and the
related effective operators which are obtained when the symmetry 
$SO(10)\times \Delta (48)\times U(1)$ is broken and heavy fermion pairs 
are integrated out: 
\begin{eqnarray*} 
W_{33} & = & \lambda_{3}16_{3}\eta_{X}\eta_{A}10_{1}\eta_{A}\eta_{X}16_{3}  
\nonumber \\
W_{32} & = & \lambda_{2}16_{3}\eta_{X}\eta_{A}\left(\frac{A_{z}}{A_{X}}\right)
10_{1}\left(\frac{A_{z}}{A_{X}}\right)\eta_{A}16_{2}  \nonumber \\
W_{22} & = &  \lambda_{2}16_{2}\eta_{A}\left(\frac{A_{u}}{A_{X}}\right) 
10_{1}\left(\frac{A_{u}}{A_{X}}\right)\eta_{A}16_{2}e^{i\phi}  \\ 
W_{12} & = & \lambda_{1}16_{1}[ \left(\frac{v_{5}}{\bar{M}_{P}}\right)^{2}
\eta'_{A}\ 10_{1}\  \eta'_{A} \nonumber \\
 & + &\left(\frac{v_{10}}{\bar{M}_{P}}\right)^{2} \eta_{A} 
\left(\frac{A_{u}}{A_{X}}\right)10_{1}\left(\frac{A_{z}}{A_{X}}\right) 
\eta_{A} ]16_{2}    \nonumber  
\end{eqnarray*}
where $\lambda_{i} = \lambda_{H}^{0}r_{i}$, $\eta_{A} = (v_{10}/A_{X})^{n+1}$, 
$\eta'_{A} = (v_{10}/A_{X})^{n-3}$.
The factor $\eta_{X} = 1/\sqrt{1 + 2\eta_{A}^{2}}$
arises from mixing, and provides a factor of $1/\sqrt{3}$ for the up-type 
quark. It remains almost unity for the down-type quark and charged lepton
as well as neutrino due to the suppression of large 
Clebsch factors in the second
term of the square root. The relative phase (or sign) between  the
two terms in the operator $W_{12}$ has been fixed.
The three directions of symmetry breaking have been chosen as
 $<A_{X}>= 2v_{10}\  diag. (1,\ 1,\ 1,\ 1,\ 1)\otimes \tau_{2}$, 
$<A_{z}> = 2v_{5}\  diag. (-\frac{1}{3},\ -\frac{1}{3},\ -\frac{1}{3},\ 
-1,\ -1)\otimes \tau_{2}$, $<A_{u}>= v_{5}/\sqrt{3}\  
diag. (2,\ 2,\ 2,\ 1,\  1)\otimes \tau_{2}$.
The resulting Clebsch  factors are
$w_{u}=w_{d}=w_{e}=w_{\nu} =1$, 
$x_{u}= 5/9$, $x_{d}= 7/27$, $x_{e}=-1/3$, $x_{\nu} = 1/5$
$y_{u}=0$, $y_{d}=y_{e}/3=2/27$, $y_{\nu} = 4/225$,  
$z_{u}=1$, $z_{d}=z_{e}= -27$, $z_{\nu} = -15^3 = -3375$, 
$z'_u = 1-5/9 = 4/9$, $z'_d = z_d + 7/729 \simeq z_{d}$,
$z'_{e} = z_{e} - 1/81 \simeq z_{e}$,   
$z'_{\nu} = z_{\nu} + 1/15^{3} \simeq z_{\nu}$.  

 An adjoint {\bf 45} $A_{X}$ and a 16-dimensional representation 
Higgs field $\Phi$ ($\bar{\Phi}$) 
are needed for breaking SO(10) down to SU(5). 
Another two adjoint 45s $A_{z}$ and $A_{u}$ are needed to break SU(5) further 
down to the standard model 
$SU(3)_{c} \times SU_{L}(2) \times U(1)_{Y}$.  
From the Yukawa coupling matrices given above ,
the 13 parameters in the SM can be determined by only four parameters: a
universal coupling constant $\lambda_{H}$ and three parameters: 
$\epsilon_{G}$, $\epsilon_{P}$ and $\tan \beta = v_2/v_1 $. 

The neutrino masses and mixings cannot be uniquely determined as they rely on
the choice of the heavy Majorana neutrino mass matrix.
 The following texture structure with zeros is found to be interesting 
for the present model
\[ M_{N}^{G} = M_{R}
\left( \begin{array}{ccc} 
0  &  0 &   \frac{1}{2}z_{N}\epsilon_{P}^{2} e^{i\delta_{\nu}}  
\\
0  &  y_{N}  & 0 \\
\frac{1}{2}z_{N}\epsilon^{2}_{P} 
e^{i\delta_{\nu}}   & 0 &  w_{N}\epsilon_{P}^{4} 
\end{array} \right) \]
 The corresponding effective operators are  
\begin{eqnarray*}  
W_{13}^{N} & = & \lambda_{1}^{N} 16_{1} (\frac{A_{z}}{v_{5}}) 
(\frac{\bar{\Phi}}{v_{10}})(\frac{\bar{\Phi}}{v_{10}}) 
(\frac{A_{u}}{v_{5}}) 16_{3}\  e^{i\delta_{\nu}} \nonumber  \\
W_{22}^{N} & = & \lambda_{2}^{N}16_{2}(\frac{A_{z}}{A_{X}}) 
(\frac{\bar{\Phi}}{v_{10}})(\frac{\bar{\Phi}}{v_{10}}) (\frac{A_{z}}{A_{X}}) 
16_{2}  \nonumber  \\  
W_{33}^{N} & = & \lambda_{3}^{N} 16_{3}(\frac{A_{u}}{v_{5}})^{2} 
(\frac{A_{z}}{v_{5}}) (\frac{\bar{\Phi}}{v_{10}})
(\frac{\bar{\Phi}}{v_{10}}) (\frac{A_{u}}{v_{5}})^{2} 
16_{3}  \nonumber 
\end{eqnarray*}
with $M_{R} = \lambda_{H}v_{10}^{2}\epsilon_{P}^{4}  
\epsilon_{G}^{2}/\bar{M}_{P}$,  $\lambda_{2}^{N} =\lambda_{H} v_{10}
\epsilon_{P}^{4}/\bar{M}_{P}$, $\lambda_{1}^{N} = \lambda_{2}^{N}
\epsilon_{P}^{2}\epsilon_{G}^{2}$ and $\lambda_{3}^{N} = \lambda_{1}^{N}
\epsilon_{P}^{2}$.
It is then not difficult to read off the Clebsch factors  
$y_{N} = 9/25$, $z_{N}= 4$ and $w_{N} = 256/27$. The CP phase $\delta_{\nu}$ is
assumed to be maximal $\delta_{\nu}=\pi /2$.

In obtaining physical masses and mixings, 
renormalization group (RG) effects should be taken into account.
The initial conditions of the RG evolution are set at the GUT scale since all
the Yukawa couplings of the quarks and leptons are generated at the GUT scale.
As most Yukawa couplings in the present model are much smaller than the top 
quark Yukawa coupling $\lambda_{t}^{G} \sim 1$, in a good approximation, we
will only keep top quark Yukawa coupling terms in the RG equations and 
neglect all other Yukawa coupling terms.
 The RG evolution  will be described by
three kinds of  scaling factors. 
$\eta_{F}$ ($F=U,D,E,N$)  and $R_{t}$
 arise from running the Yukawa parameters from the GUT scale 
down to the SUSY breaking scale $M_{S}$ which is chosen to be close to
the top quark mass, i.e., $M_{S} \simeq m_{t}\simeq 170$ GeV. 
They are defined by $\eta_{F}(M_{S}) = \prod_{i=1}^{3}
\left(\frac{\alpha_{i}(M_{G})}{\alpha_{i}(M_{S})}\right)^{c_{i}^{F}/2b_{i}}$ 
$(F=U, D, E, N)$ with $c_{i}^{U} = (\frac{13}{15}, 3, \frac{16}{3})$, 
$c_{i}^{D} = (\frac{7}{15}, 3, \frac{16}{3})$ ,  
$c_{i}^{E} = (\frac{27}{15}, 3, 0)$, $c_{i}^{N} = (\frac{9}{25}, 3, 0)$,   
$b_{i} = (\frac{33}{5}, 1, -3)$,  
and $ R_{t}^{-1} = exp[-\int_{\ln M_{S}}^{\ln M_{G}} 
(\frac{\lambda_{t}(t)}{4\pi})^{2} dt ]  
=[1 + (\lambda_{t}^{G})^{2} K_{t}]^{-1/12}$, 
where $K_{t} = \frac{3 I(M_{S})}{4\pi^{2}}$ with 
$I(M_{S}) = \int_{\ln M_{S}}^{\ln M_{G}} \eta_{U}^{2}(t) dt $ 
with $M_{S} \simeq m_{t} = 170$GeV. Other RG scaling factors are derived by
running Yukawa couplings below $M_{S}$. $m_{i}(m_{i}) = \eta_{i} \  
m_{i} (M_{S})$ for $(i = c,b )$ and 
 $m_{i}(1GeV) = \eta_{i}\  m_{i} (M_{S})$ for ($i = u,d,s$).
The physical top quark mass is given by 
$M_{t} = m_{t}(m_{t}) \left(1 +
\frac{4}{3}\frac{\alpha_{s}(m_{t})}{\pi}\right)$. 
 The scaling factor $R_{t}$ or coupling 
$\lambda_{t}^{G} = \frac{1}{\sqrt{K_{t}}} \frac{\sqrt{1 -
R_{t}^{-12}}}{R_{t}^{-6}}$  is 
determined by the mass ratio of the bottom quark and $\tau$ lepton.
$\tan \beta$ is fixed by the $\tau$ lepton mass via 
$\cos \beta = \frac{m_{\tau} \sqrt{2}}{\eta_{E} \eta_{\tau} v 
\lambda_{\tau}^{G}} $.  
In numerical predictions we take $\alpha^{-1}(M_Z) = 127.9 $, $s^{2}(M_Z) 
= 0.2319$, $M_Z = 91.187$ GeV, 
$\alpha_{1}^{-1}(m_t) = 58.59$, 
$\alpha_{2}^{-1}(m_t) = 30.02$ and 
$\alpha_{1}^{-1}(M_G) = \alpha_{2}^{-1}(M_G) = \alpha_{3}^{-1}(M_G)
 \simeq 24 $ with $M_{G} \sim 2 \times 10^{16}$ GeV.  For $\alpha_{s}(M_{Z}) = 
0.113$, the RG scaling factors have values ($\eta_{u,d,s}$, $\eta_{c}$,
$\eta_{b}$, $\eta_{e,\mu,\tau}$, $\eta_{U}$, $\eta_{D}/\eta_{E}\equiv
\eta_{D/E}$, $\eta_{E}$, $\eta_{N}$) = (2.20, 2.00, 1.49, 1.02, 3.33, 2.06,
1.58, 1.41).
The corresponding predictions on fermion masses and mixings thus obtained are 
found to be remarkable.  Our numerical
predictions for $\alpha_{s} (M_{Z}) = 0.113$ are given in table 1 with   four 
input parameters. Where $B_{K}$ and $f_{B} \sqrt{B}$ 
in table 1 are two important hadronic
parameters and extracted from $K^{0}-\bar{K}^{0}$ and  $B^{0}-\bar{B}^{0}$
mixing parameters $\varepsilon_{K}$ and $x_{d}$. $Re(\varepsilon'/\varepsilon)$
is the direct CP-violating parameter in kaon decays, where large 
uncertanties mainly arise from the hadronic matrix elements. $\alpha$, 
$\beta$ and $\gamma$ are three angles of the unitarity triangle in the 
Cabibbo-Kobayashi-Maskawa (CKM) matrix. 
$J_{CP}$ is the rephase-invariant CP-violating quantity. The light neutrino
masses and mixings are obtained via see-saw mechanism $M_{\nu} =
\Gamma^{G}_{\nu} (M_{N}^{G})^{-1}(\Gamma_{\nu}^{G})^{\dagger}
v_{2}^{2}/(2R_{t}^{-6}\eta_{N}^{2})$. The nine predictions on $|V_{us}|$,  
$|V_{ub}|/|V_{cb}|$, $|V_{td}|/|V_{ts}|$, $m_{d}/m_{s}$, $|V_{\nu_{\mu}e}|$,
$|V_{\nu_{\mu}\tau}|$, $|V_{\nu_{e}\tau}|$ as well as
$m_{\nu_{e}}/m_{\nu_{\mu}}$ and $m_{\nu_{\mu}}/m_{\nu_{\tau}}$
are obtained from nine RG-independent relations and given solely by the Clebsch
factors and the three charged lepton masses.
  
\begin{table*}[hbt]
\setlength{\tabcolsep}{1.5pc}
\newlength{\digitwidth} \settowidth{\digitwidth}{\rm 0}
\catcode`?=\active \def?{\kern\digitwidth}
\caption{Output observables and model parameters and 
their predicted values with $\alpha_{s}(M_{Z}) = 0.113$ and 
input parameters: $m_{e} = 0.511$ eV, 
$m_{\mu} = 105.66$ MeV, $m_{\tau} = 1.777$ GeV, and $m_{b}(m_{b}) = 4.25$ GeV.}
\label{tab:predictions}
\begin{tabular*}{\textwidth}{@{}l@{\extracolsep{\fill}}rrlr} 
\hline
 Output    &  Output    &  Data\cite{DATA} & 
 Output    &  Output     \\ \hline 
$M_{t}$\ [GeV]  &  182   &  $180 \pm 15 $  &  $J_{CP}/10^{-5}$ & $2.68 $  \\
$m_{c}(m_{c})$\ [GeV]  &  1.27   & $1.27 \pm 0.05$  & 
 $\alpha$ & $86.28^{\circ}$ \\ 
$m_{u}$(1GeV)\ [MeV]  &  4.31   &  $4.75 \pm 1.65$ & 
$\beta$ & $22.11^{\circ}$ \\
$m_{s}$(1GeV)\ [MeV]  &  156.5  &  $165\pm 65$  &  
$\gamma$ & $71.61^{\circ}$  \\
$m_{d}$(1GeV) \ [MeV]  &  6.26 & $8.5 \pm 3.0$ & 
$m_{\nu_{\tau}}$ [eV]  & $ 2.4515$    \\
$|V_{us}|=\lambda $ & 0.22 & $0.221 \pm 0.003$ & 
$m_{\nu_{\mu}}$ [eV]  & $2.4485$   \\
$\frac{|V_{ub}|}{|V_{cb}|} $ & 0.083 & 
$0.08 \pm 0.03$ & $m_{\nu_{e}}$ [eV]/$10^{-3}$ & $ 1.27$   \\
$\frac{|V_{td}|}{|V_{ts}|}$ & 0.209 & 
$0.24 \pm 0.11$ & $m_{\nu_{s}}$ [eV]/$10^{-3}$  & $ 2.8 $  \\
 $|V_{cb}|=A\lambda^{2}$ & 0.0393  &  $0.039 \pm 0.005 $ &
  $|V_{\nu_{\mu}e}| $ &  -0.049  \\
$\lambda_{t}^{G}$  & 1.30  & - &  $|V_{\nu_{e}\tau}| $ &  0.000  \\
$\tan \beta = v_{2}/v_{1}$ & 2.33 & - &   $|V_{\nu_{\tau}e}| $ & -0.049   \\
$\epsilon_{G}$ &  $0.2987$ & - & 
$|V_{\nu_{\mu}\tau}| $ &  -0.707  \\
$\epsilon_{P}$  & $0.0101$ & - & 
$|V_{\nu_{e}s}|$ & $ 0.038 $ \\
$B_{K}$ & 0.90 &  $0.82 \pm 0.10$ & $M_{N_{1}}$ [GeV] & $\sim 333$  \\
$f_{B}\sqrt{B}$ [MeV] & 207  & $200 \pm 70 $ &
$M_{N_{2}}$ [GeV]/$10^{6}$ & $1.63$  \\
 Re($\varepsilon'/\varepsilon)/10^{-3} $ & $1.4 \pm 1.0$ &  
$1.5 \pm 0.8 $ & $M_{N_{3}}$ [GeV] & $ 333$  
  \\  \hline

\end{tabular*}
\end{table*}

From the results in table 1, we observe the following: 
{\bf 1}. a $\nu_{\mu} (\bar{\nu}_{\mu}) \rightarrow \nu_{\tau} 
(\bar{\nu}_{\tau})$ long-wave length oscillation with 
$\Delta m_{\mu\tau}^{2} = m_{\nu_{\tau}}^{2} - m_{\nu_{\mu}}^{2} \simeq 
1.5 \times 10^{-2} eV^{2}$ and 
$\sin^{2}2\theta_{\mu\tau} \simeq 0.987$
could explain the atmospheric neutrino 
deficit\cite{ATMO};
{\bf 2}. Two massive neutrinos $\nu_{\mu}$ and $\nu_{\tau}$ with 
$m_{\nu_{\mu}} \simeq m_{\nu_{\tau}}  \simeq 2.45$  eV 
fall in the range required by  
possible hot dark matter\cite{DARK};
{\bf 3} a short wave-length oscillation with 
$\Delta m_{e\mu}^{2} = m_{\nu_{\mu}}^{2} - m_{\nu_{e}}^{2} 
\simeq 6\  eV^{2}$ an $\sin^{2}2\theta_{e\mu} \simeq 1.0 \times 10^{-2}$ 
is consistent with the LSND experiment\cite{LSND}. 
{\bf 4}. $(\nu_{\mu} - \nu_{\tau})$ oscillation will be beyond the reach
of CHORUS/NOMAD and E803. However, $(\nu_{e} - \nu_{\tau})$ oscillation
may become interesting as a short wave-length oscillation with 
$\Delta m_{e\tau}^{2} = m_{\nu_{\tau}}^{2} - m_{\nu_{e}}^{2} 
\simeq 6\  eV^{2}$ and 
$\sin^{2}2\theta_{e\tau} \simeq 1.0 \times 10^{-2}$;  
{\bf 5}. Majorana neutrino  allows neutrinoless double beta decay
$(\beta \beta_{0\nu})\cite{DBETA}$. The decay rate is found to be
$ \Gamma_{\beta\beta} \simeq 1.0 \times 10^{-61}$ GeV which is below to the
present upper limit;
{\bf 6}. solar neutrino deficit has to be explained by oscillation 
between $\nu_{e}$ and  a sterile neutrino $\nu_{s}$
\cite{STERILE}(singlet of SU(2)$\times$ U(1), or 
singlet of SO(10) in the GUT SO(10) model). 
Masses and mixings of the triplet sterile neutrinos can be chosen 
by introducing an additional  singlet scalar with VEV $v_{s}\simeq 336$ GeV.
They are found to be $m_{\nu_{s}} = \lambda_{H} v_{s}^{2}/v_{10} 
\simeq 2.8 \times 10^{-3} $eV
and $\sin\theta_{es} \simeq m_{\nu_{L}\nu_{s}}/m_{\nu_{s}} 
= v_{2}\epsilon_{P}/(2v_{s}\epsilon_{G}^{2})\simeq 3.8 
\times 10^{-2}$. The resulting parameters 
$\Delta m_{es}^{2} = m_{\nu_{s}}^{2} - m_{\nu_{e}}^{2} 
\simeq 6.2\times 10^{-6}\  eV^{2}$ and 
$\sin^{2}2\theta_{es} \simeq 5.8 \times 10^{-3}$;  
 are consistent with the values \cite{STERILE} 
obtained from fitting the experimental data.

 It is amazing that nature has allowed us to 
make predictions on fermion masses and mixings 
in terms of a single Yukawa coupling constant and 
three ratios determined by the structure of the physical 
vacuum and  understand the low energy physics 
from the GUT scale physics. It has also suggested that nature favors 
 maximal spontaneous CP violation. 
It is expected that more precise measurements from CP violation, neutrino
oscillation and various low
energy experiments in the near future could provide a good test on the 
present model and guide us to a more fundamental model.

{\bf ACKNOWLEDGEMENTS}:\   YLW would like to thank professor 
R. Mohapatra for a kind invitation to him to present this work 
at the 4th SUSY96 conference held at University of Maryland, 
May 29- June 1, 1996.

\end{document}